\documentclass[final]{svjour3}

\usepackage[english]{babel}

\usepackage{mathrsfs}
\usepackage{amssymb,amsmath}
\usepackage{amsfonts}
\usepackage{latexsym}

\usepackage{graphicx}

\usepackage{float}

\def\lb{\label}

\newcommand{\er}[1]{\textrm{(\ref{#1})}}

\usepackage{color}
\def\rev#1{{#1}}

\def\eeq{\relax}
\def\beq#1#2\eeq{\begin{equation}\label{#1}#2\end{equation}}
\def\bal#1#2\eal{\begin{align}\label{#1}#2\end{align}}
\def\bse#1#2\ese{\begin{subequations}\label{#1}#2\end{subequations}}

\begin{document}



\def\a{\alpha}  \def\cA{{\mathcal A}}     \def\bA{{\bf A}}  \def\mA{{\mathscr A}}
\def\b{\beta}   \def\cB{{\mathcal B}}     \def\bB{{\bf B}}  \def\mB{{\mathscr B}}
\def\g{\gamma}  \def\cC{{\mathcal C}}     \def\bC{{\bf C}}  \def\mC{{\mathscr C}}
\def\G{\Gamma}  \def\cD{{\mathcal D}}     \def\bD{{\bf D}}  \def\mD{{\mathscr D}}
\def\d{\delta}  \def\cE{{\mathcal E}}     \def\bE{{\bf E}}  \def\mE{{\mathscr E}}
\def\D{\Delta}  \def\cF{{\mathcal F}}     \def\bF{{\bf F}}  \def\mF{{\mathscr F}}
\def\c{\chi}    \def\cG{{\mathcal G}}     \def\bG{{\bf G}}  \def\mG{{\mathscr G}}
\def\z{\zeta}   \def\cH{{\mathcal H}}     \def\bH{{\bf H}}  \def\mH{{\mathscr H}}
\def\e{\eta}    \def\cI{{\mathcal I}}     \def\bI{{\bf I}}  \def\mI{{\mathscr I}}
\def\p{\psi}    \def\cJ{{\mathcal J}}     \def\bJ{{\bf J}}  \def\mJ{{\mathscr J}}
\def\vT{\Theta} \def\cK{{\mathcal K}}     \def\bK{{\bf K}}  \def\mK{{\mathscr K}}
\def\k{\kappa}  \def\cL{{\mathcal L}}     \def\bL{{\bf L}}  \def\mL{{\mathscr L}}
\def\l{\lambda} \def\cM{{\mathcal M}}     \def\bM{{\bf M}}  \def\mM{{\mathscr M}}
\def\L{\Lambda} \def\cN{{\mathcal N}}     \def\bN{{\bf N}}  \def\mN{{\mathscr N}}
\def\m{\mu}     \def\cO{{\mathcal O}}     \def\bO{{\bf O}}  \def\mO{{\mathscr O}}
\def\n{\nu}     \def\cP{{\mathcal P}}     \def\bP{{\bf P}}  \def\mP{{\mathscr P}}
\def\r{\rho}    \def\cQ{{\mathcal Q}}     \def\bQ{{\bf Q}}  \def\mQ{{\mathscr Q}}
\def\s{\sigma}  \def\cR{{\mathcal R}}     \def\bR{{\bf R}}  \def\mR{{\mathscr R}}
\def\cS{{\mathcal S}}     \def\bS{{\bf S}}  \def\mS{{\mathscr S}}
\def\t{\tau}    \def\cT{{\mathcal T}}     \def\bT{{\bf T}}  \def\mT{{\mathscr T}}
\def\f{\phi}    \def\cU{{\mathcal U}}     \def\bU{{\bf U}}  \def\mU{{\mathscr U}}
\def\F{\Phi}    \def\cV{{\mathcal V}}     \def\bV{{\bf V}}  \def\mV{{\mathscr V}}
\def\P{\Psi}    \def\cW{{\mathcal W}}     \def\bW{{\bf W}}  \def\mW{{\mathscr W}}
\def\o{\omega}  \def\cX{{\mathcal X}}     \def\bX{{\bf X}}  \def\mX{{\mathscr X}}
\def\x{\xi}     \def\cY{{\mathcal Y}}     \def\bY{{\bf Y}}  \def\mY{{\mathscr Y}}
\def\X{\Xi}     \def\cZ{{\mathcal Z}}     \def\bZ{{\bf Z}}  \def\mZ{{\mathscr Z}}
\def\O{\Omega}

\newcommand{\gA}{\mathfrak{A}}
\newcommand{\gB}{\mathfrak{B}}
\newcommand{\gC}{\mathfrak{C}}
\newcommand{\gD}{\mathfrak{D}}
\newcommand{\gE}{\mathfrak{E}}
\newcommand{\gF}{\mathfrak{F}}
\newcommand{\gG}{\mathfrak{G}}
\newcommand{\gH}{\mathfrak{H}}
\newcommand{\gI}{\mathfrak{I}}
\newcommand{\gJ}{\mathfrak{J}}
\newcommand{\gK}{\mathfrak{K}}
\newcommand{\gL}{\mathfrak{L}}
\newcommand{\gM}{\mathfrak{M}}
\newcommand{\gN}{\mathfrak{N}}
\newcommand{\gO}{\mathfrak{O}}
\newcommand{\gP}{\mathfrak{P}}
\newcommand{\gQ}{\mathfrak{Q}}
\newcommand{\gR}{\mathfrak{R}}
\newcommand{\gS}{\mathfrak{S}}
\newcommand{\gT}{\mathfrak{T}}
\newcommand{\gU}{\mathfrak{U}}
\newcommand{\gV}{\mathfrak{V}}
\newcommand{\gW}{\mathfrak{W}}
\newcommand{\gX}{\mathfrak{X}}
\newcommand{\gY}{\mathfrak{Y}}
\newcommand{\gZ}{\mathfrak{Z}}

\def\ve{\varepsilon}   \def\vt{\vartheta}    \def\vp{\varphi}    \def\vk{\varkappa}

\def\Z{{\mathbb Z}}    \def\R{{\mathbb R}}   \def\C{{\mathbb C}}    \def\K{{\mathbb K}}
\def\T{{\mathbb T}}    \def\N{{\mathbb N}}   \def\dD{{\mathbb D}}


\def\la{\leftarrow}              \def\ra{\rightarrow}            \def\Ra{\Rightarrow}
\def\ua{\uparrow}                \def\da{\downarrow}
\def\lra{\leftrightarrow}        \def\Lra{\Leftrightarrow}


\def\lt{\biggl}                  \def\rt{\biggr}
\def\ol{\overline}               \def\wt{\widetilde}
\def\no{\noindent}


\let\ge\geqslant                 \let\le\leqslant
\def\lan{\langle}                \def\ran{\rangle}
\def\/{\over}                    \def\iy{\infty}
\def\sm{\setminus}               \def\es{\emptyset}
\def\ss{\subset}                 \def\ts{\times}
\def\pa{\partial}                \def\os{\oplus}
\def\om{\ominus}                 \def\ev{\equiv}
\def\iint{\int\!\!\!\int}        \def\iintt{\mathop{\int\!\!\int\!\!\dots\!\!\int}\limits}
\def\el2{\ell^{\,2}}             \def\1{1\!\!1}
\def\sh{\sharp}
\def\wh{\widehat}
\def\bs{\backslash}

%
\def\all{\mathop{\mathrm{all}}\nolimits}
\def\Area{\mathop{\mathrm{Area}}\nolimits}
\def\arg{\mathop{\mathrm{arg}}\nolimits}
\def\const{\mathop{\mathrm{const}}\nolimits}
\def\det{\mathop{\mathrm{det}}\nolimits}
\def\diag{\mathop{\mathrm{diag}}\nolimits}
\def\diam{\mathop{\mathrm{diam}}\nolimits}
\def\dim{\mathop{\mathrm{dim}}\nolimits}
\def\dist{\mathop{\mathrm{dist}}\nolimits}
\def\Im{\mathop{\mathrm{Im}}\nolimits}
\def\Iso{\mathop{\mathrm{Iso}}\nolimits}
\def\Ker{\mathop{\mathrm{Ker}}\nolimits}
\def\Lip{\mathop{\mathrm{Lip}}\nolimits}
\def\rank{\mathop{\mathrm{rank}}\limits}
\def\Ran{\mathop{\mathrm{Ran}}\nolimits}
\def\Re{\mathop{\mathrm{Re}}\nolimits}
\def\Res{\mathop{\mathrm{Res}}\nolimits}
\def\res{\mathop{\mathrm{res}}\limits}
\def\sign{\mathop{\mathrm{sign}}\nolimits}
\def\supp{\mathop{\mathrm{supp}}\nolimits}
\def\Tr{\mathop{\mathrm{Tr}}\nolimits}
\def\BBox{\hspace{1mm}\vrule height6pt width5.5pt depth0pt \hspace{6pt}}
\def\where{\mathop{\mathrm{where}}\nolimits}
\def\as{\mathop{\mathrm{as}}\nolimits}


\newcommand\nh[2]{\widehat{#1}\vphantom{#1}^{(#2)}}
\def\dia{\diamond}

\def\Oplus{\bigoplus\nolimits}



\def\qqq{\qquad}
\def\qq{\quad}
\let\ge\geqslant
\let\le\leqslant
\let\geq\geqslant
\let\leq\leqslant
\newcommand{\ca}{\begin{cases}}
\newcommand{\ac}{\end{cases}}
\newcommand{\ma}{\begin{pmatrix}}
\newcommand{\am}{\end{pmatrix}}
\renewcommand{\[}{\begin{equation}}
\renewcommand{\]}{\end{equation}}
\def\eq{\begin{equation}}
\def\qe{\end{equation}}
\def\[{\begin{equation}}
\def\bu{\bullet}

\newcommand{\bk}{\mbox{\boldmath  $\kappa$ }}
\newcommand{\bm}{\mbox{\boldmath  $\mu$ }}
\newcommand{\ba}{\mbox{\boldmath  $\alpha$ }}
\newcommand{\bbe}{\mbox{\boldmath  $\beta$ }}

\title{Converging bounds for the effective shear speed in 2D phononic crystals}
\author{A.A.Kutsenko \and A.L.Shuvalov \and A.N.Norris}
\institute{A.A.Kutsenko \and A.L.Shuvalov \at Universit\'{e} de
Bordeaux, Institut de M\'{e}canique et d'Ing\'{e}nierie de Bordeaux,
UMR 5295, Talence 33405, France \\ \email{{\color{blue}aak@nxt.ru}}
\and A.N.Norris \at Mechanical and Aerospace Engineering, Rutgers
University, Piscataway, NJ 08854, USA}

\def\Wr{\mathop{\rm Wr}\nolimits}
\def\BBox{\hspace{1mm}\vrule height6pt width5.5pt depth0pt \hspace{6pt}}

\def\Diag{\mathop{\rm Diag}\nolimits}

\date{\today}


\maketitle

\begin{abstract}
Calculation of the effective quasistatic shear speed $c$ in 2D solid
phononic crystals is analyzed. The plane-wave expansion (PWE) and
the monodromy-matrix (MM) methods are considered. For each method,
the stepwise sequence of upper and lower bounds is obtained which
monotonically converges to the exact value of $c$. It is proved that
the two-sided MM bounds of $c$ are tighter and their convergence to
$c$ is uniformly faster than that of the PWE bounds. Examples of the
PWE and MM bounds of effective speed versus concentration of
high-contrast inclusions are demonstrated.
\end{abstract}

\section{Introduction}

Recent progress in fabrication of periodic composite materials has
intensified interest in their effective elastic properties. One of
these parameters is the quasistatic limit of the shear speed $c$
defined by the ratio of effective shear to averaged density.
The effective speed $c$ may vary significantly at small changes of  the filling
fraction in high-contrast phononic crystals, thus evaluation of $c$
needs to be reliable and accurate. Except for certain model cases
(see an example in Appendix \ref{A1}), the effective speed does not admit a
closed-form, i.e. exact, value and has to be calculated
numerically by one of the known series or iterative schemes. Despite
the broad application of these methods, a quantitative analysis of their
convergence is lacking. As a result, it is not evident how
to pinpoint the deviation of numerically obtained $c$ from its
actual value and thus to describe the accuracy of its calculation.

Addressing this fundamental question, the present paper provides
explicit majorant and minorant stepwise sequences which
monotonically converge to the exact effective speed $c$ in a 2D
cubic lattice with isotropic shear properties. Such sequences of
two-sided bounds of $c$ are obtained for the two key methods of the
effective speed calculation: one is the broadly used method of
plane-wave expansion (PWE) \cite{KAG}; the other is the recently
proposed method of monodromy matrix (MM) \cite{KSNP,KSN}. It is
shown that, for any fixed step $N$, the pair of MM bounds lies in
between the PWE bounds. Hence the MM bounds enable a more
accurate capture of the exact $c$ and have a faster convergence to $c$ as $%
N\rightarrow \infty $ than the PWE bounds.

The paper is organised as follows. Two equivalent analytical
definitions of the effective speed $c$ are given in \S2. The main
results on the PWE and MM sequences of two-sided bounds of $c$ are
formulated in \S3. These results are illustrated for several
examples of two- and three-phase periodic solid composites in \S4
where the PWE and MM bounds are calculated and plotted at a fixed
step $N$ as functions of filling fraction. The proofs of the
theorems of \S3 are given in \S5. The conclusions follow in \S6.
Some auxiliary remarks are provided in the Appendix.

\section{Background}

We consider the \rev{time harmonic} wave equation
\rev{for shear horizontal (SH) motion}
\[\lb{001}
 -\nabla\cdot\m\nabla v=\r\o^2v,
\]
where $\nabla=(\pa_i)_{i=1}^2$, $\pa_i=\pa/\pa x_i$ and $\cdot$ is a
scalar product. The shear coefficient $\m$ and the density $\r$ are
real positive ${\bf1}$-periodic functions on a 2D \rev{square} unit cell:
\[\lb{002}
 \m,\r({\bf x}+{\bf e}_i)={\m},\r({\bf x}),\ \
 \forall {\bf
 x}\in\R^2;\ \ \ {\bf e}_i=\rev{(\d_{i1}, \d_{i2})},\ \ i=1,2,
\]
where $\d$ is the Kronecker symbol. Assume $v$ in the Floquet form
$v=e^{i {\bf k}\cdot{\bf x}}u$ with ${\bf 1}$-periodic function $u$
and the Floquet vector ${\bf k}\in\R^2$. Then the operator $\cC
v\ev-\nabla\cdot\m\nabla v$ of \er{001} can be cast as
\[\lb{dirint}
 \cC({\bf k})u=-(\nabla+i{\bf k})\cdot\m(\nabla+i{\bf
k})u.
\]
For any fixed ${\bf k}$, the operator $\cC({\bf k})$ has purely
discrete spectrum $\o_1^2({\bf k})\le\o^2_2({\bf k})\le...$, where
$\o_n({\bf k})$ are called Floquet branches. Note that $\o_1({\bf
0})=0$ is an eigenvalue of $\cC({\bf 0})$ with multiplicity $1$ and
the corresponding eigenfunction is $u_1\ev1$. The effective speed is
introduced as
\[\lb{009}
 c(\boldsymbol{\k})=\lim_{k\to0}\frac{\o_1({\bf k})}{k},\ \ {\rm where}\ \ {\bf
 k}=k\boldsymbol{\k},\ \ \|\boldsymbol{\k}\|=1.
\]
Expanding \er{dirint} as
\[\lb{006}
 \cC({\bf k})=\cC_0+k\cC_1+k^2\cC_2,\ \
 \cC_0 u=-\nabla\cdot{\m}\nabla u,
\ \
 \cC_1
 u=-i\boldsymbol{\k}\cdot{\m}\nabla u-i\nabla\cdot{\m}\boldsymbol{\k}u,\ \
 \cC_2u={\m}u
\]
and applying \rev{regular}  perturbation theory to \er{001} (see Lemma
\ref{pert}) defines $c(\boldsymbol{\k})$ by the formula
\[\lb{010}
 c^2(\boldsymbol{\k})=\frac{\langle{\m}\rangle-
 (\cC_1\cC_0^{-1}\cC_1u_1,u_1)}{\langle\r\rangle},
\]
where $\langle\cdot\rangle=\int_{[0,1]^2}\cdot d{\bf x}$ and
$(u,v)=\langle u\ol{v}\rangle$ denotes the standard inner product in
$L^2([0,1]^2)$. Though \er{010} is an explicit definition of $c$, it
still  requires calculation of the inverse of the operator $\cC_0$,
which in general has no exact closed form except for some special
cases  (see an example in Appendix \ref{A1}).

There exists another explicit representation for $c(\boldsymbol{\k})$ in
terms of the  monodromy matrix \cite{KSNP,KSN}. For $\boldsymbol{\k}$
along the principal direction (e.g. ${\bf e}_1$), this
representation yields
\[\lb{mon}
 c^2({\bf e}_1)=\frac1{\langle\r\rangle}\ma 0 \\ 1 \am\cdot(\cM-\cI)^{-1}\ma 1 \\ 0 \am,\ \
 \cM=\wh{\int}_0^1(\cI+\cQ dx_1),\ \ \cQ=\ma 0 & \m^{-1} \\ -\pa_2\m\pa_2 & 0
 \am,
\]
where $\cI$ is the identity operator and $\wh{\int}$ is the multiplicative integral (see Appendix \ref{A2}). However \er{mon} is also not a
closed-form solution.

We do not discuss the domain of definition of $\wh\int$ of the
infinite-dimensional operator $\cQ$ since we will actually use
$\wh\int$ of only finite-dimension matrices (see \er{022}), in which
case $\wh\int$ is well defined.

Hereafter for brevity we restrict consideration to the typical case
of the function $\m({\bf x})$ satisfying cubic symmetry $\m(\mathbf{R}{\bf
x})=\m({\bf x})$, where $\mathbf{R}$ is a matrix of rotation by
$\frac{\pi}2$. In this case the effective speed does not depend on
$\boldsymbol{\k}$, i.e. $c(\boldsymbol{\k})=\const$, and \er{010} can be rewritten
as
\[\lb{meff}
 c^2=\frac{\m_{\rm eff}(\m)}{\langle\r\rangle}\ \ {\rm with}\ \ \m_{\rm
 eff}(\m)=\langle{\m}\rangle-
 (\cC_1\cC_0^{-1}\cC_1u_1,u_1),
\]
where the effective shear coefficient $\m_{\rm eff}(\m)$ is a
functional depending on the function $\m$ only. Assumption of cubic
symmetry also allows us to use the identity
\[\lb{012}
 \m_{\rm eff}(\m)=1/\m_{\rm eff}(\m^{-1}),
\]
which is proved in \cite{NK} and in \cite{JKO} by variational methods. 
This identity is
instrumental in the following derivations, where we  will show that
the approximations of $\m_{\rm eff}(\m)$ and of $1/\m_{\rm
eff}(\m^{-1})$ lead to the upper and lower bounds of $\m_{\rm eff}$,
respectively.

\section{Two-sided bounds of $c$}
Due to \er{meff}$_1$, it suffices to obtain bounds of $\m_{\rm
eff}$.
\subsection{PWE method}
This method is based on using the formula \er{010} with $\cC_0$,
$\cC_1$ restricted to the space of the first $(2N+1)^2$ simple harmonics
$e^{2\pi i{\bf g}\cdot{\bf x}}$. Denote the Fourier coefficients of
the function $\m$ by $\wh \m({\bf g})$, i.e.
\[\lb{012a}
 \m({\bf x})=\sum_{{\bf g}\in\Z^2}\wh{ \m}({\bf g})e^{2\pi i{\bf g}\cdot{\bf
 x}},
\]
and introduce the $(2N+1)^2\ts(2N+1)^2$ matrix and $(2N+1)^2$-vector
\[\lb{013}
 \cC_{NN,0}\ev(\wh{{\m}}({\bf g}-{\bf g}'){\bf g}\cdot{\bf g}')_{|g'_i|,|g_i|\le
 N},\ \
 {\bf f}_{NN}\ev(\wh{{\m}}({\bf g})g_1)_{|g_i|\le N},
\]
where ${\bf g}=(g_1,g_2)^{\top}\in\Z^2$. Define the functionals
$\m_{NN}$ and $\wt\m_{NN}$ by
\[\lb{016}
 {{\m}}_{NN}(\m)=\langle{\m}\rangle-{\bf f}_{NN}\cdot\cC_{NN,0}^{-1}{\bf
 f}_{NN},\ \ \wt\m_{NN}(\m)=1/\m_{NN}(\m^{-1}),
\]
where the definition of $\m_{NN}(\m^{-1})$ in \er{016}$_2$ implies
substitution of $\m^{-1}$ instead of $\m$ in \er{012a}-\er{016}$_1$.
Note that $\cC_{NN,0}^{-1}$ does not exist, since it has null vector
$(\d_{{\bf g}{\bf 0}})$, but $\cC_{NN,0}^{-1}{\bf f}_{NN}$ exists as
a preimage of ${\bf f}_{NN}$ under the action of $\cC_{NN,0}$ (this
preimage is not unique but the scalar product in \er{016} is, since
the scalar product ${\bf f}_{NN}\cdot(\d_{{\bf g}{\bf 0}})=0$). Let us
formulate the first result.
\begin{theorem}\lb{T1}
The sequence ${\m}_{NN}$ monotonically decreases to ${\m}_{\rm
eff}$, the sequence $\wt{{\m}}_{NN}$ monotonically increases to
$\m_{\rm eff}$, i.e.
\[\lb{018}
 {\m}_{NN}\searrow{\m}_{\rm eff},\ \ \ \wt{{\m}}_{NN}\nearrow{\m}_{\rm eff},\ \ \ N\to\iy.
\]
\end{theorem}
Note that \er{018} with $N=0$ is the Voigt-Reuss bound
$\langle{\m}^{-1}\rangle^{-1}\le{\m}_{\rm
eff}\le\langle{\m}\rangle$, see \cite{JKO}.


\subsection{MM method}
This method is based on using the formula \er{mon} with $\cQ(x_1)$
restricted to the space of the first $2N+1$ simple harmonics $e^{2\pi
inx_2}$. Denote the Fourier coefficients of $\m(x_1,x_2)$ in $x_2$
by $\wh \m_n(x_1)$, i.e.
\[\lb{020}
 \m(x_1,x_2)=\sum_{n\in\Z}\wh \m_n(x_1)e^{2\pi inx_2},
\]
and introduce the $(2N+1)\ts(2N+1)$ matrices
\[\lb{021}
 \wh{\boldsymbol{\m}}_N\ev\wh{\boldsymbol{\m}}_N(x_1)=(\wh \m_{n-m}(x_1))_{n,m=-N}^N,
\ \
 \boldsymbol{\pa}_N=2\pi\diag(n)_{-N}^{N}.
\]
Define the $(4N+2)\ts(4N+2)$  matrix ${\bf Q}_N$ and the corresponding
multiplicative integral by
\[\lb{022}
 {\bf Q}_N=\ma {\bf 0} & \wh{\boldsymbol{\m}}_N^{-1} \\
               \boldsymbol{\pa}_N\wh{\boldsymbol{\m}}_N\boldsymbol{\pa}_N &
               {\bf 0}      \am,\ \ {\bf M}_N=\wh{\int_0^1}({\bf I}+{\bf
               Q}_Ndx_1),
\]
where ${\bf I}$ is the identity matrix. Define functionals $\m_N$ and
$\wt\m_N$ by
\[\lb{024} \m_N(\m)={\bf
w}_2\cdot({\bf M}_N-{\bf I})^{-1}{\bf w}_1,\ \
\wt\m_N(\m)=1/\m_N(\m^{-1}),
\]
where
\[\lb{024aa}
 {\bf w}_1=\ma
 {\bf e}_{(N)} \\ {\bf 0} \am,\ \ {\bf w}_2=\ma
 {\bf 0} \\ {\bf e}_{(N)} \am,\ \ {\bf e}_{(N)}=(\d_{0n})_{n=-N}^N.
\]
Note that $({\bf M}_N-{\bf I})^{-1}$ does not exist but $({\bf
M}_N-{\bf I})^{-1}{\bf w}_1$ exists as the preimage of ${\bf w}_1$
(this preimage is not unique but the scalar product in \er{024} is).
We now  formulate the main result.
\begin{theorem}\lb{T2}
i) The sequence ${\m}_{N}$ monotonically decreases to ${\m}_{\rm
eff}$, the sequence $\wt{{\m}}_{N}$ monotonically increases to
${\m}_{\rm eff}$, i.e.
\[\lb{028}
 {\m}_{N}\searrow{\m}_{\rm eff},\ \ \
 \wt{{\m}}_{N}\nearrow{\m}_{\rm eff},\ \ \ N\to\iy.
\]
ii) Moreover,
\[\lb{029}
 \wt{{\m}}_{NN}\le\wt{{\m}}_{N}\le{\m}_{\rm eff}\le{{\m}}_{N}\le{{\m}}_{NN},\
 \ \ \forall N,
\]
and hence the convergence in \er{028} is faster than in \er{018}.
\end{theorem}
Note that \er{028} with $N=0$ yields the known inequality $
\langle\langle{\m}^{-1}\rangle_2^{-1}\rangle_1\le{\m}_{\rm eff}
 \le\langle\langle{\m}\rangle_2^{-1}\rangle_1^{-1}$, see \cite{JKO}.

The MM bounds \er{024} admit a simpler form if the function $\m$ is
even in at least one argument. Denote the multiplicative integral
over half of the period as
\[\lb{030}
 {\bf M}_{N,\frac12}=\wh{\int_0^\frac12}({\bf I}+{\bf Q}_Ndx_1)
\]
and let ${\bf m}_N$ be the upper right $(2N+1)\ts(2N+1)$ block of
${\bf M}_{N,\frac12}$. Taking \er{022} and \er{030} with the
function $\m^{-1}$ defines $\wt{\bf m}_N$, i.e. $\wt{\bf
m}_N(\m)$=${\bf m}_N(\m^{-1})$. The following result holds true.
\begin{theorem} \lb{T4}
Suppose that $\m(-x_1,x_2)=\m(x_1,x_2)$ for all $x_1,x_2$. Then
$\m_N$ and $\wt\m_N$ which appear in \er{028} can also be defined by
\[\lb{031}
 \m_N=\frac12{\bf e}_{(N)}\cdot{\bf m}_{N}^{-1}{\bf e}_{(N)},\ \
 \wt\m_N=2({\bf e}_{(N)}\cdot\wt{\bf m}_{N}^{-1}{\bf e}_{(N)})^{-1},
\]
where ${\bf e}_{(N)}$ is given by \er{024aa}.
\end{theorem}

In conclusion let us summarize the results in terms of the effective
speed $c=\sqrt{\m_{\rm eff}/\langle\r\rangle}$. Introduce the PWE
and MM bounds of $c$ as, respectively,
\[\lb{cn}
 c_{NN}=\sqrt{\frac{\m_{NN}}{\langle\r\rangle}},\ \
 \wt c_{NN}=\sqrt{\frac{\wt\m_{NN}}{\langle\r\rangle}};\ \ \
 c_{N}=\sqrt{\frac{\m_{N}}{\langle\r\rangle}},\ \
 \wt c_{N}=\sqrt{\frac{\wt\m_{N}}{\langle\r\rangle}},
\]
where $\m_{NN}$, $\wt\m_{NN}$ are given by \er{016} and $\m_N$,
$\wt\m_N$ are given by \er{024} or, for the even $\m$, by \er{031}.
According to \er{018}, \er{028} and \er{029},
\[\lb{ecn1}
 c_{NN}\searrow c,\ \ \wt c_{NN}\nearrow c;\ \ \ c_{N}\searrow c,\ \ \wt
 c_{N}\nearrow c\ \ {\rm and}\ \
 \wt c_{NN}\le \wt c_{N}\le c\le c_{N} \le c_{NN}.
\]
Note that $c\approx c_{NN}$ is the result of \cite{KAG} and that
$c\approx c_N$ was exemplified in \cite{KSN}.

\section{Examples}
We provide several examples of the PWE and MM bounds \er{cn},
\er{ecn1} of the effective speed $c$ in high-contrast two- and
three-phase lattices. Their profiles admit application of \er{031}.
The results are presented for different $N$ as functions of filling
fraction $f$. The PWE and MM bounds are displayed by dashed and
solid lines, respectively (colored online).

It is observed that MM bounds provide a noticeably sharper
estimation of the exact effective speed than the PWE bounds. For the
two-phase lattices one of the PWE bounds is close to the exact
effective speed (see Figs. \ref{fig1}b and \ref{fig2}b), but this is
no longer so for three-phase lattices (see Figs. \ref{fig3} and
\ref{fig4}).

Regarding high-contrast two-component materials it is also
noteworthy that the upper bounds ($c_N$, $c_{NN}$) and lower bounds
($\wt c_N$, $\wt c_{NN}$) give better approximations of $c$ in the
case of the stiff matrix/soft inclusion and of the soft matrix/stiff
inclusion, respectively, see Figs. \ref{fig1},\ref{fig2}.

Fast convergence of the MM bounds shown in Fig. \ref{fig2}b confirms
the conclusion of \cite{KSN} that the exact dependence $c(f)$ for
densely packed stiff inclusions is more accurately described by the
MM curve $c_N(f)$ with a steep trend at $f\to1$ than by the PWE
curve $c_{NN}(f)$ with inflexion (the latter PWE curve was used as a
numerical benchmark in \cite{KSNP}).

\begin{figure}[h]
\begin{minipage}[h]{0.45\linewidth}
\center{\includegraphics[width=1\linewidth]{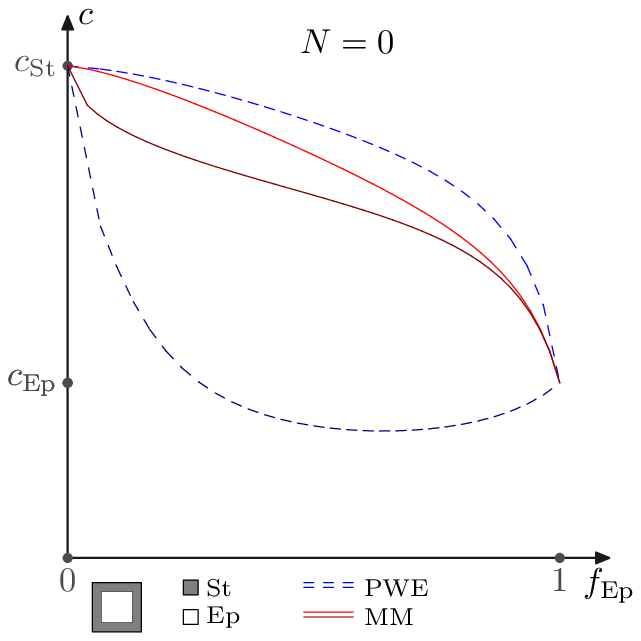} \\ a)}
\end{minipage}
\hfill
\begin{minipage}[h]{0.45\linewidth}
\center{\includegraphics[width=1\linewidth]{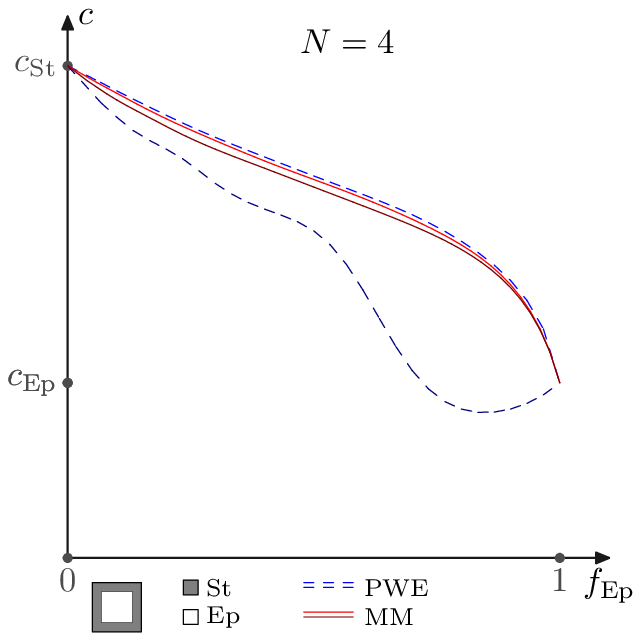} \\ b)}
\end{minipage}
\caption{PWE ($c_{NN}$, $\wt c_{NN}$) and MM ($c_N$, $\wt c_{N}$)
bounds for Steel/Epoxy lattice of nested squares: a) $N=0$, b)
$N=4$.} \label{fig1}
\end{figure}

\begin{figure}[h]
\begin{minipage}[h]{0.45\linewidth}
\center{\includegraphics[width=1\linewidth]{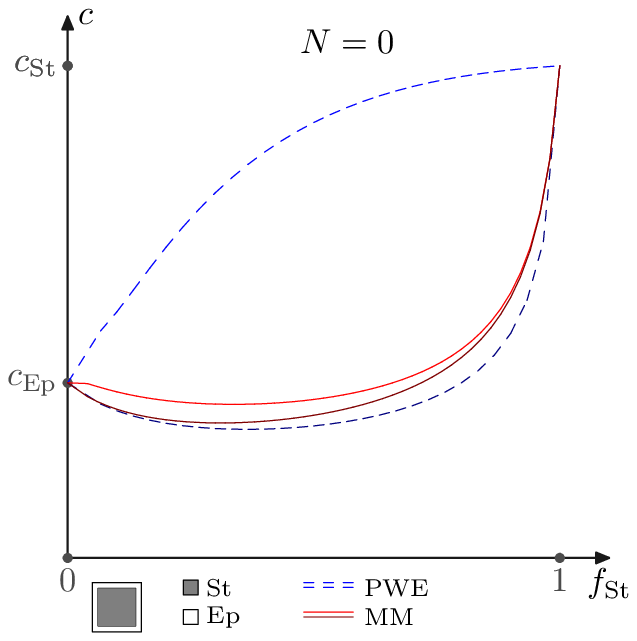} \\ a)}
\end{minipage}
\hfill
\begin{minipage}[h]{0.45\linewidth}
\center{\includegraphics[width=1\linewidth]{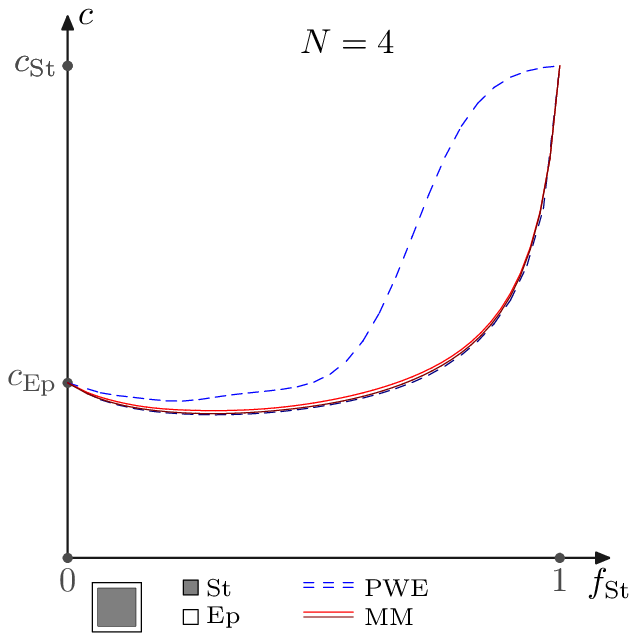} \\ b)}
\end{minipage}
\caption{PWE ($c_{NN}$, $\wt c_{NN}$) and MM ($c_N$, $\wt c_{N}$)
bounds for  Epoxy/Steel lattices of nested squares: a) $N=0$, b)
$N=4$.} \label{fig2}
\end{figure}

\begin{figure}[h]
\begin{minipage}[h]{0.49\linewidth}
\center{\includegraphics[width=1\linewidth]{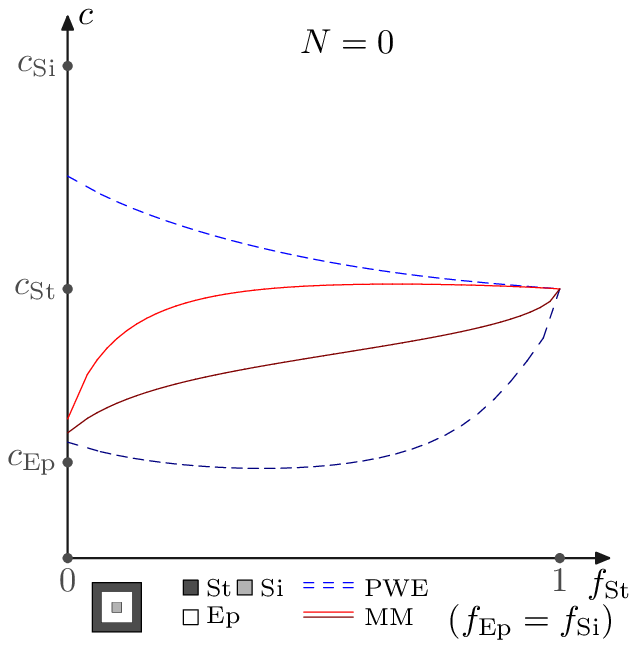} \\ a)}
\end{minipage}
\hfill
\begin{minipage}[h]{0.49\linewidth}
\center{\includegraphics[width=1\linewidth]{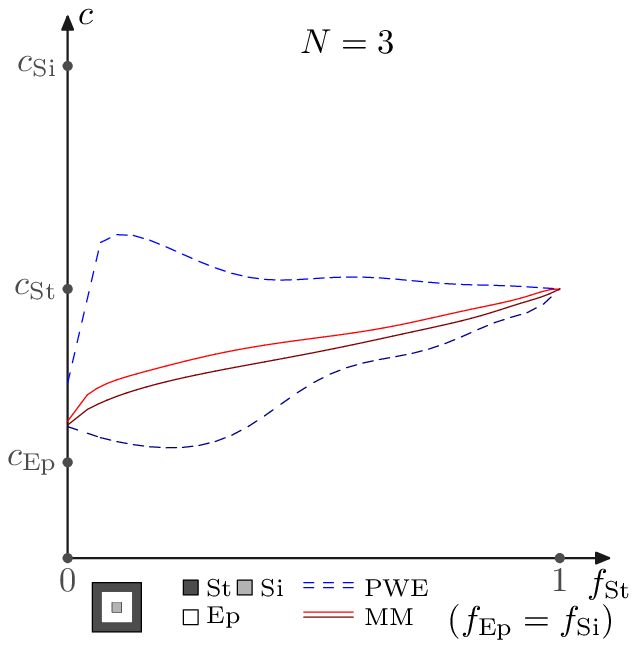} \\ b)}
\end{minipage}
\vfill
\begin{minipage}[h]{0.49\linewidth}
\center{\includegraphics[width=1\linewidth]{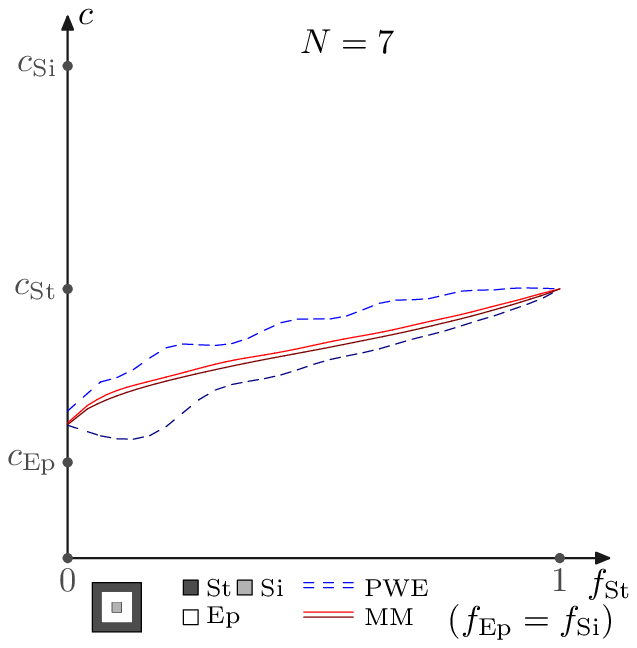} \\ c)}
\end{minipage}
\hfill
\begin{minipage}[h]{0.49\linewidth}
\center{\includegraphics[width=1\linewidth]{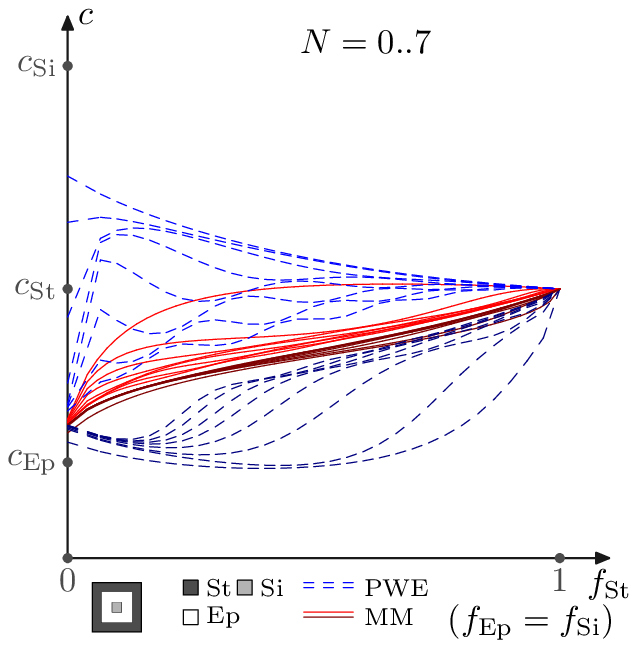} \\ d)}
\end{minipage}
\caption{PWE ($c_{NN}$, $\wt c_{NN}$) and MM ($c_N$, $\wt c_{N}$)
bounds for  Steel/Epoxy/Silicon lattice of nested squares: a) $N=0$,
b) $N=3$, c) $N=7$, d) $N=0,...,7$.} \label{fig3}
\end{figure}

\begin{figure}[h]
\begin{minipage}[h]{0.49\linewidth}
\center{\includegraphics[width=1\linewidth]{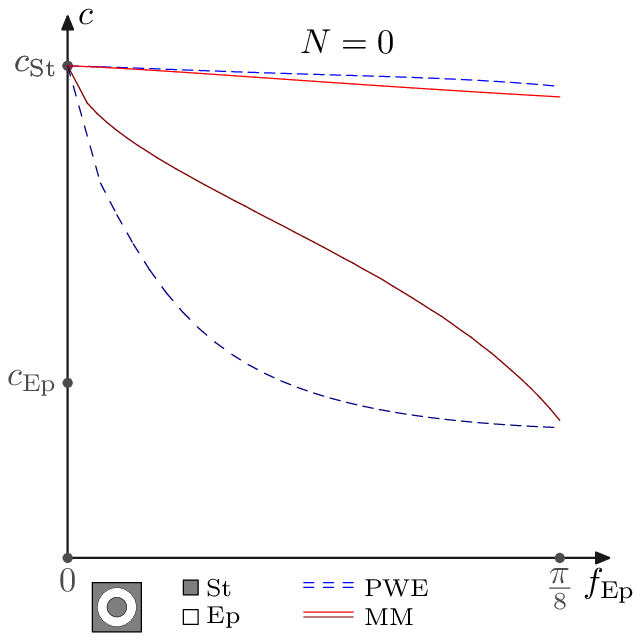} \\ a)}
\end{minipage}
\hfill
\begin{minipage}[h]{0.49\linewidth}
\center{\includegraphics[width=1\linewidth]{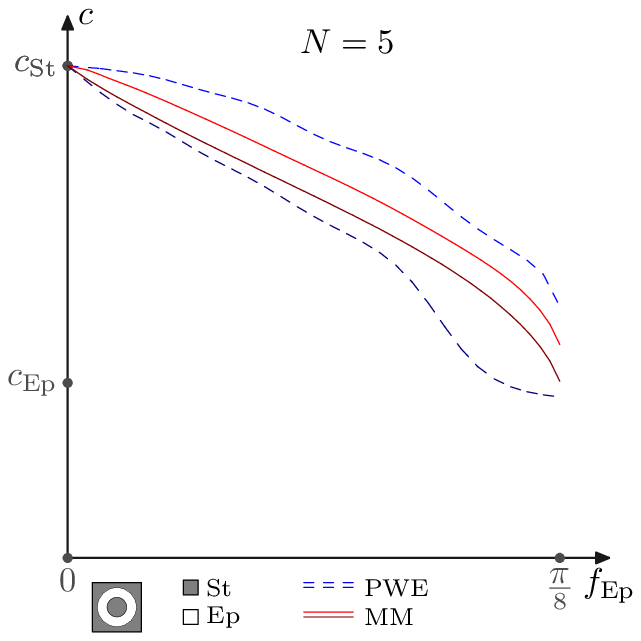} \\ b)}
\end{minipage}
\caption{PWE ($c_{NN}$, $\wt c_{NN}$) and MM ($c_N$, $\wt c_{N}$)
bounds for  Steel/Epoxy/Steel lattice of nested circles: a) $N=0$,
b) $N=5$.} \label{fig4}
\end{figure}

\section{Proof of the main results}

\begin{lemma} \lb{pert} Consider the eigenvalue problem
\[\lb{l01}
 {\bf A}{\bf u}=\l{\bf B}{\bf u}\ \ {\rm with}\ \ {\bf A}={\bf A}_0+k{\bf A}_1+k^2{\bf
 A}_2,
\]
where ${\bf A}$, ${\bf B}$ ($\det{\bf B}\ne0$) are self-adjoint
matrices and $k$ is a small real parameter. Suppose that $\l=0$ for
$k=0$ is a simple eigenvalue of ${\bf A}$ with normalized
eigenvector ${\bf u}_0$ (${\bf u}_0\cdot{\bf u}_0=1$). Then
\[\lb{l02}
 \l(k)=k\l_1+k^2\l_2+O(k^3),\ \ {\bf u}(k)={\bf u}_0+k{\bf u}_1+{\bf
 O}(k^2),
\]
\[\lb{l03}
 \l_1=\frac{{\bf A}_1{\bf u}_0\cdot{\bf u}_0}{{\bf B}{\bf u}_0\cdot{\bf u}_0},\ \
 {\bf u}_1={\bf A}_0^{-1}(\l_1{\bf B}-{\bf A}_1){\bf
 u}_0,\ \
 \l_2=\frac{{\bf A}_2{\bf u}_0\cdot{\bf u}_0-{\bf A}_0{\bf u}_1\cdot{\bf
 u}_1}{{\bf B}{\bf u}_0\cdot{\bf u}_0}.
\]
\end{lemma}
{\it Proof.} Substituting expansions \er{l03} into \er{l01} and
equating the terms with the same power of $k$ we obtain
\[\lb{l04}
 {\bf A}_0{\bf u}_1+{\bf A}_1{\bf u}_0=\l_1{\bf B}{\bf u}_0,
\]
\[\lb{l05}
 {\bf A}_0{\bf u}_2+{\bf A}_1{\bf u}_1+{\bf A}_2{\bf u}_0=\l_2{\bf B}{\bf
 u}_0+\l_1{\bf B}{\bf u}_1.
\]
Scalar multiplying both sides in \er{l04}, \er{l05} by ${\bf u}_0$,
using ${\bf A}_0{\bf u}_0={\bf 0}$ and self-adjointness of ${\bf
A}_i$, ${\bf B}$ we obtain \er{l03}. \BBox

\subsection{Proof of Theorem \ref{T1}} According to \er{009} and \er{meff}, the effective shear coefficient
$\m_{\rm eff}$ given by \er{meff}$_2$ can be defined as
\[\lb{p021}
 \m_{\rm eff}=\lim_{k\to0}\frac{\o_1^2}{k^2},
\]
where $\o_1$ is a minimal eigenvalue of the eigenproblem \er{001}
with $\r=1$, i.e. of
\[\lb{101}
 \cC({\bf k})u=\o^2({\bf k})u.
\]
Introduce the subspace $\cL_{NN}$ of $L^2([0,1]^2)$,
\[\lb{100}
 \cL_{NN}=\cL(\{e^{2\pi i(g_1x_1+g_2x_2)}:\ \ |g_1|,|g_2|\le N\}),
\]
where $\cL(\cdot)$ means the linear span of the set. Denote the
corresponding projector $\cP_{NN}:L^2([0,1]^2)\to\cL_{NN}$. Consider
the equation
\[\lb{102}
 \cC_{NN}({\bf k})u=\o^2_{NN}({\bf k})u
\]
with
\[\lb{103}
 \cC_{NN}({\bf k}):\cL_{NN}\to\cL_{NN},\ \ \ \cC_{NN}({\bf
 k})\ev\cP_{NN}\cC({\bf k}).
\]
The operator $\cC_{NN}$ can be represented as a finite matrix
\[\lb{104}
 \cC_{NN}({\bf k})=(\wh{{\m}}({\bf g}-{\bf g}')(2\pi{\bf g}+{\bf k})\cdot(2\pi{\bf g}'+{\bf
 k}))_{|g_i'|,|g_i|\le N},
\]
where $\wh{{\m}}$ are Fourier coefficients for ${\m}$ (see
\er{012a}). Note that the minimal eigenvalue $\o_{1,NN}({\bf k})$ of
$\cC_{NN}({\bf k})$ is greater than the minimal eigenvalue
$\o_1({\bf k})$ of $\cC({\bf k})$, since
\[\lb{105}
 \o_1({\bf k})=\inf_{u\in \cH}\frac{(\cC({\bf k}) u,u)}{(u,u)}\le
 \inf_{u\in\cL_{NN}}\frac{(\cC({\bf k}) u,u)}{(u,u)}=
 \inf_{u\in\cL_{NN}}\frac{(\cC_{NN}({\bf k})
 u,u)}{(u,u)}=\o_{1,NN}({\bf k}),
\]
where $\cH$ is a Sobolev space. Also $\o_{1,NN}\searrow\o_1$ for
$N\to\iy$, since $\cL_{N'N'}\ss\cL_{NN}$ for $N'\le N$ and
$\cup_N\cL_{NN}$ is dense in $\cH$. Denote the limit
\[\lb{106}
 b_{NN}(\boldsymbol{\k})=\lim_{k\to0}\frac{\o^2_{1,NN}({\bf k})}{k^2}\ \ \ ({\bf
 k}=k\boldsymbol{\k},\ \ \|\boldsymbol{\k}\|=1).
\]
By \er{105}, $b_{NN}$ is greater than $\m_{\rm eff}$ and
$b_{NN}\searrow\m_{\rm eff}$ for $N\to\iy$. Taking ${\bf k}=k{\bf
e}_1$ ($\boldsymbol{\k}={\bf e}_1$) in \er{104} leads to
\begin{multline}\lb{asc}
 \cC_{NN}=\cC_{NN,0}+k\cC_{NN,1}+k^2\cC_{NN,2}\ \ {\rm where}
 \\
 \cC_{NN,1}=2\pi((g_1+g'_1)\wh{{\m}}({\bf g}-{\bf g}'))_{|g_i|\le
 N},\ \ \cC_{NN,2}=(\wh{{\m}}({\bf g}-{\bf g}'))_{|g_i|\le
 N}
\end{multline}
and $\cC_{NN,0}$ is given by \er{013}. Applying Lemma \ref{pert} to
${\bf A}=\cC_{NN}$, ${\bf B}={\bf I}$ with ${\bf u}_0=(\d_{{\bf
g}{\bf 0}})_{|g_i|\le N}$ and $\l=\o_{1,NN}^2=\l_1k+b_{NN}k^2+...$
yields $\l_1=0$ and $b_{NN}=\m_{NN}$, where $\m_{NN}$ is given by
\er{016}$_1$. Since $b_{NN}\searrow\m_{\rm eff}$ (see above), we
conclude that $\m_{NN}\searrow\m_{\rm eff}$. Applying the same steps
to $\m^{-1}$ and using \er{012}, \er{016}$_2$ we obtain
${\wt\m}_{NN}\nearrow\m_{\rm eff}$. Thus \er{018} is proved. \BBox

\subsection{Proof of Theorem \ref{T2}}

As in the previous section, we proceed from equation \er{102}.
Introduce the subspace $\cL_{N}$ of $\cH_{\bf k}=e^{i{\bf
k}\cdot{\bf x}}\cH$,
\[\lb{300}
 \cL_{N}=\cL(\{e^{i\bf{k}\cdot{\bf x}}e^{2\pi i(g_1x_1+g_2x_2)}:\ \ g_1\in\Z,\ |g_2|\le
 N\}),
\]
and the corresponding projector $\cP_{N}:\cH_{\bf k}\to\cL_{N}$.
Consider the equation
\[\lb{302}
 \cC_{N}({\bf k})u=\o^2_{N}({\bf k})u
\]
with
\[\lb{303}
 \cC_{N}({\bf k}):\cL_{N}\to\cL_{N},\ \ \ \cC_{N}({\bf
 k})\ev\cP_{N}\cC({\bf k}).
\]
Suppose that ${\bf k}=\ma k_1 & 0\am^{\top}$, i.e. $k_2=0$. Let
$\o_{1,N}$ be the minimal eigenvalue  in \er{302}, and denote the
limit
\[\lb{t01}
 b_N=\lim_{k_1\to0}\frac{\o^2_{1,N}(k_1{\bf e}_1)}{k^2_1}.
\]
Repeating arguments from \er{105} to the end of \S5.1 and using the
fact that $e^{i{\bf k}\cdot{\bf x}}\cL_{NN}\ss \cL_N$ (see \er{100},
\er{300}) we obtain
\[\lb{t02}
 b_N\searrow\m_{\rm eff},\ \ (\m_{NN}=)b_{NN}\ge b_N.
\]
In order to complete the proof, we need to show that $b_N=\m_N$. The
operator $\cC_{N}$ can be represented as a 1D vector differential
operator
\[\lb{304}
 \cC_{N}=-\pa_1\wh{\boldsymbol{\m}}_N\pa_1+\boldsymbol{\pa}_N\wh{\boldsymbol{\m}}_N\boldsymbol{\pa}_N,
\]
where the notations \er{021} are used. Equation \er{302} can be
rewritten in the form
\begin{equation}\lb{305}
 \pa_1\boldsymbol{\eta}=\wt{\bf Q}_N\boldsymbol{\eta} \ \  {\rm with} \quad
\left\{
\begin{aligned}
 \boldsymbol{\eta}=\ma {\bf u}_N &
 \wh{\boldsymbol{\m}}_N\pa_1{\bf u}_N
 \am^{\top},\ \ {\bf u}_N(x_1)&= \big(\wh u_n(x_1)\big)_{n=-N}^{N},
 \\
\quad  \wt{\bf Q}_N=\ma {\bf 0} & \wh{\boldsymbol{\m}}_N^{-1} \\
 \boldsymbol{\pa}_N\wh{\boldsymbol{\m}}_N\boldsymbol{\pa}_N-\o_N^2{\bf I} &
 {\bf 0}\am.
\end{aligned}
\right.
\end{equation}
The solution of \er{305} has the following form
\[\lb{307}
 \boldsymbol{\eta}(x_1)=\wt{\bf M}_N[x_1,0]\boldsymbol{\eta}(0)\ \ {\rm with}\ \
 \wt{\bf M}_N[\b,\a]=\wh{\int_{\a}^{\b}}({\bf I}+\wt{\bf Q}_N dx_1).
\]
Taking $x_1=1$ in \er{307} and noting that
$\boldsymbol{\eta}=e^{ik_1x_1}\boldsymbol{\xi}$ with periodic $\boldsymbol{\xi}$ (see
\er{300}) we obtain
\[\lb{309}
 e^{ik_1}\boldsymbol{\xi}(0)=\wt{\bf M}_N[1,0]\boldsymbol{\xi}(0).
\]
In order to find $b_N$ \er{t01} we need the asymptotics of each term in
\er{309}. Using \er{t01}, we expand  $\wt{\bf Q}_N$ \er{305} as
\[\lb{311}
 \wt{\bf Q}_N={\bf Q}_N+k_1^2b_N{\bf Q}_{2,N}+...\ \ {\rm with}\ \
 {\bf Q}_{2,N}=\ma {\bf 0} & {\bf 0} \\
 -{\bf I} &
 {\bf 0}\am
\]
and substitute it into \er{307}$_2$ to obtain
\begin{multline}\lb{315}
 \wt{\bf M}_N[1,0]={\bf M}_N+k_1^2b_N{\bf M}_{2,N}+...\ \ {\rm with}\\
 {\bf M}_{2,N}=\int_0^1{\bf M}_N[1,x]{\bf Q}_{2,N}(x){\bf
 M}_N[x,0]dx,\ \
 {\bf M}_N[\b,\a]=\wh{\int_{\a}^{\b}}({\bf I}+{\bf Q}_Ndx_1),
\end{multline}
where ${\bf Q}_N$ and ${\bf M}_N={\bf M}_N[1,0]$ are given in
\er{022}. Note that
\[\lb{320}
 {\bf Q}_N{\bf w}_1={\bf 0},\ \ {\bf w}_2^*{\bf Q}_N={\bf 0}
\]
for ${\bf w}_i$ defined in \er{024aa}, since  ${\bf
e}_{(N)}^*\boldsymbol{\pa}_N=\boldsymbol{\pa}_N{\bf e}_{(N)}={\bf 0}$. Combining
\er{320} with \er{022}$_2$ yields
\[\lb{321}
 {\bf M}_N[\b,\a]{\bf w}_1={\bf w}_1,\ \ {\bf w}_2^*{\bf M}_N[\b,\a]={\bf
 w}_2^{*},\ \ \forall\, \a,\b.
\]
Hence, by \er{315} and \er{321}, the vector $\boldsymbol{\xi}(0)$ in
\er{309} satisfies
\[\lb{322}
 \boldsymbol{\xi}(0)={\bf w}_1+k_1\boldsymbol{\xi}_1+k_1^2\boldsymbol{\xi}_2+...
\]
with unknown $\boldsymbol{\xi}_1$, $\boldsymbol{\xi}_2, \ldots$. Substituting
\er{322}, \er{315} with $e^{ik_1}=1+ik_1-\frac 12 {k_1^2} +...$ into
\er{309} and equating terms with the same power of $k_1$ yields
\[\lb{323}
 \boldsymbol{\xi}_1=i({\bf M}_N-{\bf I})^{-1}{\bf w}_1,
\]
\[\lb{324}
 {\bf M}_N\boldsymbol{\xi}_2+b_N{\bf M}_{2,N}{\bf w}_1=\boldsymbol{\xi}_2+i\boldsymbol{\xi}_1-\frac{1}{2}{\bf
 w}_1.
\]
Multiplying \er{324} by the vector ${\bf w}_2^*$ and using \er{321}
along with ${\bf w}_2^*{\bf w}_1=0$, we obtain
\[\lb{325}
 b_N={\bf w}_2^*({\bf M}_N-{\bf I})^{-1}{\bf w}_1,
\]
which coincides with $\m_N$ in \er{024}. Thus \er{t02} yields
\er{028}, \er{029} for the upper bound $\m_N$. The proof for the
lower bound $\wt\m_N$ is similar. \BBox

\subsection{Proof of Theorem \ref{T4}}  Taking \er{325} with $b_N=\m_N$, applying the chain
rule, and using \er{321}, we obtain
\[\lb{t001}
 \m_N={\bf w}_2^*({\bf M}_N-{\bf I})^{-1}{\bf w}_1=
 {\bf w}_2^*({\bf M}_N[\tfrac 12,0]-{\bf M}_N[\tfrac 12,1])^{-1}{\bf w}_1.
\]
The definition \er{315} of ${\bf M}_N[\b,\a]$ and the ${\bf 1}$-periodicity
of $\m$ with symmetry $\m(x_1,x_2)=\m(-x_1,x_2)$ give us
\[\lb{t002}
 {\bf M}_N[\tfrac 12,0]=\wh{\int_0^{\frac12}}({\bf I}+{\bf Q}_Ndx_1),\ \
 {\bf M}_N[\tfrac 12,1]=\wh{\int_0^{\frac12}}({\bf I}-{\bf Q}_Ndx_1).
\]
Due to \er{t002} we get
\[\lb{t003}
 {\bf M}_N[\tfrac 12,0]=\ma{\bf a}_1 & {\bf a}_2 \\ {\bf a}_3 & {\bf a}_4\am,\
 \ {\bf M}_N[\tfrac 12,1]=\ma{\bf a}_1 & -{\bf a}_2 \\ -{\bf a}_3 & {\bf
 a}_4\am,
\]
since  blocks of ${\bf Q}_N$ on the diagonal are zero matrices, see (\ref{022}). Equalities
\er{t001}, \er{t003} and ${\bf M}_{N,\frac12}={\bf M}_N[\tfrac 12,0]$ lead
to \er{031}$_1$. The proof of \er{031}$_2$ is similar. \BBox

\section{Conclusion}

The PWE and MM bounds of the effective speed $c$ have been
presented. It was shown that the MM bounds $c_N$, $\wt c_N$ are more
accurate than the PWE bounds $c_{NN}$, $\wt c_{NN}$. In fact even for
not so large $N$ it is often sufficient to use only one MM bound
$c_N$ or $\wt c_{N}$ to obtain a good enough approximation of $c$.

Moreover, numerical implementation of the MM scheme requires less
computation time per  step than the PWE method, since the former
needs to calculate an exponent of $(4N+2)\ts(4N+2)$ matrix and to
solve a system of $(4N+2)$ linear equations whereas the latter needs
to solve a system of $(2N+1)^2$ linear equations.

The results of the paper apply to other types of scalar waves
described by the governing equations similar to \er{001}, such as
acoustic waves in fluids, and electromagnetic waves.

\renewcommand{\appendix}{
  \setcounter{section}{0}\renewcommand{\thesection}{\Alph{section}}
    \setcounter{equation}{0}
  \renewcommand{\theequation}{\thesection.\arabic{equation}}
    }
\appendix
\section{Appendix}

\subsection{Example of a closed form $c(\boldsymbol{\k})$} \label{A1}

Suppose that $\m=\m_1(x_1)\m_2(x_2)$. Then $c(\boldsymbol{\k})$ admits a
closed-form representation
\[\lb{010a}
  c^2(\boldsymbol{\k})=\frac1{\langle\r\rangle}\boldsymbol{\k}^{\top}\ma \langle\m_2\rangle_2\langle\m_1^{-1}\rangle_1^{-1} & 0 \\
                         0 & \langle\m_1\rangle_1\langle\m_2^{-1}\rangle_2^{-1}
                         \am\boldsymbol{\k},
\]
where $\langle\cdot\rangle_i=\int_0^1\cdot dx_i$. The proof of
\er{010a} is based on the fact that the equation $\cC_0h=\cC_1u_1$
(see \er{006}) has closed-form solution $h=\cC_0^{-1}\cC_1u_1$. Let
$\boldsymbol{\k}={\bf e}_1$, then
\[\lb{p01}
 \cC_0h=\cC_1u_1\ \Rightarrow\
 -\m_2\pa_1(\m_1\pa_1h)- \m_1\pa_2(\m_2\pa_2h)=-i\m_2\pa_1\m_1.
\]
Assume the solution of \er{p01} in the form $h=h(x_1)$, then
\[\lb{p02}
 \pa_1(\m_1\pa_1h)=i\pa_1\m_1\ \Rightarrow\ \m_1\pa_1h=i\m_1+\a_1\ \Rightarrow\
 h=\a_2+ix_1+\a_1\int_0^{x_1}\m_1^{-1}dx_1,
\]
\[\lb{p03}
 h(1)=h(0)\ \Rightarrow\
 h=\a_2+ix_1-i\langle\m_1^{-1}\rangle^{-1}\int_0^{x_1}\m_1^{-1}dx_1\ \ (\a_2=\const).
\]
Substituting $h=\cC_0^{-1}\cC_1u_1$ from \er{p03} into \er{010}
gives the upper left element of the matrix in \er{010a}. Other elements
are obtained similarly. If $\m$ depends on $x_1$ only, then
\er{010a} reduces to the well-known result $\langle\r\rangle
c^2=\langle\m^{-1}\rangle^{-1}{\k_1}^2+\langle\m\rangle\k_2^2$.

\subsection{Options for calculating the multiplicative integral}\label{A2}
By definition, the multiplicative integral ${\bf
M}[\b,\a]=\wh\int_{\a}^{\b}({\bf I}+{\bf Q}dx)$ is
\[\lb{034}
 {\bf M}=\lim_{k\to\iy}\prod_{j=[\a k]}^{[\b k]}({\bf I}+(1/k){\bf
 Q}(j/k)),
\]
where $[\cdot]$ denotes integer part. This formula is
straightforward for numerical implementation. It was employed for
calculating \er{030} to obtain the MM curves for circular
inclusions, see Fig.\ref{fig4}.

Another method is to use the Peano series
\[\lb{033}
 {\bf M}={\bf I}+\int_{\a}^{\b}{\bf Q}(y_1)dy_1+\int_{\a}^{\b}\int_{\a}^{y_1}{\bf
 Q}(y_1){\bf Q}(y_2)dy_1dy_2+....
\]
It converges faster than \er{034} (at the same rate as the series for
exponent of ${\bf Q}$) but its implementation is more laborious.

If ${\bf Q}(x)$ is a piecewise constant function on $[\a,\b]$, i.e.
\begin{multline}\lb{append1}
 {\bf Q}(x)=\const_i\ {\rm for}\ x\in\D_i\ {\rm where}\\
 [\a,\b]=\cup_{i=1}^n\D_i,\ \ \D_i=[x_{i-1},x_i),\ \
 \a=x_0<x_1<...<x_n=\b,
\end{multline}
then
\[\lb{032}
 {\bf M}=\exp(|\D_n|{\bf Q}(x_{n-1}))...\exp(|\D_1|{\bf Q}(x_0)).
\]
This formula was used for calculating \er{030} to obtain the MM
curves in Figs. \ref{fig1}-\ref{fig3}.

Note in conclusion that the principal formula \er{024} involves the
resolvent $({\bf M}_N-{\bf I})^{-1}$, which can be calculated
directly (i.e. without evaluating ${\bf M}_N$) by numerical
integration of the corresponding Riccati equation. In fact using the
resolvent has some numerical advantage, because the increase of its
elements with growing $N$ is slower than the increase of elements of
${\bf M}_N$.

\begin{acknowledgements} We thank T. A. Suslina for very useful
discussions. One of the authors (AAK) acknowledges support from the
University Bordeaux 1 via the project AP\_ 2011.

\end{acknowledgements}

\end{document}